\documentclass[journal]{IEEEtran}

\usepackage{booktabs}
\usepackage{array}
\usepackage{url}
\usepackage{amsmath}
\usepackage{graphicx}
\usepackage{multirow}

\begin{document}

\title{SMSI: System Model Security Inference: Automated Threat Modeling for Cyber-Physical Systems}

\author{Ro\'Yah~Radaideh~and~Ali~Khreis%
\thanks{All authors are with the School of Electrical Engineering and Computer Science, University of Ottawa, Ottawa, ON, Canada. Course: ELG5271/CSI5388 -- AI for Cybersecurity Applications. Instructor: Prof.\ Paula Branco. Winter 2025/2026.}}

\markboth{ELG5271/CSI5388 -- AI for Cybersecurity Applications, University of Ottawa, Winter 2025/2026}{}

\maketitle

\begin{abstract}
Threat modeling for cyber-physical systems (CPS) remains a largely manual exercise.
Security analysts receive a system architecture diagram, scan vulnerability databases,
and then spend hours deciding which attacks are plausible and which controls to
implement. This project asks whether that workflow can be automated.
We present SMSI (System Model Security Inference), a hybrid neuro-symbolic pipeline
that starts from a SysML architecture model and produces a prioritized list of
NIST 800-53 security controls, with every recommendation traced back to the specific
system component, vulnerability, and attacker technique that motivated it.
The prototype has three main stages: a deterministic parser that maps system components to
software vulnerabilities via the National Vulnerability Database (NVD); a family of
retrieval and classification models that link those vulnerabilities to MITRE ATT\&CK
techniques; and a control recommender that combines the Center for Threat-Informed
Defense (CTID) ATT\&CK$\to$800-53 crosswalk with text similarity and CVSS-weighted
prioritization. 

We explore three complementary approaches for CVE-to-ATT\&CK mapping:
(1)~a supervised multi-label classifier using fine-tuned SecureBERT+, achieving
Micro~F1 of 0.757 on 105 parent technique classes;
(2)~retrieval-based dense encoders (SecureBERT, ATTACK-BERT, MiniLM) with
KEV-supervised fine-tuning, where MiniLM achieves the best MRR of 0.252 and
Hits@10 of 0.578 on a 45-query KEV test split; and
(3)~a zero-shot LLM approach using Gemma-4 26B via LM~Studio, which achieves
51.8\% Hit Rate@1 on 419 KEV CVEs-substantially outperforming all embedding
baselines on hit-rate metrics.

We validate the pipeline on a synthetic healthcare IoT gateway,
MedGateway, with nine software components yielding 199 CVEs.
For the ATT\&CK-to-NIST stage, pretrained SecureBERT achieves the highest control
retrieval scores (MRR~0.582, Hits@10~0.798), demonstrating that the curated CTID
crosswalk combined with dense embeddings provides a strong basis for automated control
recommendation.
Taken together, SMSI reduces what would otherwise require days of expert manual
analysis to an automated, fully traceable report suitable for analyst review.
\end{abstract}


\IEEEpeerreviewmaketitle

\section{Introduction}
\label{sec:intro}

\IEEEPARstart{W}{hen} a security team inherits a new cyber-physical system (say, a
hospital IoT gateway that connects infusion pumps to a cloud analytics platform),
their job is to figure out what can go wrong and what to do about it. In practice,
this means consulting the system architecture, searching the National Vulnerability
Database (NVD) for known flaws in the deployed software, figuring out which adversary
techniques those flaws enable, and then mapping those techniques to the relevant
NIST 800-53 controls. Each step is intellectually non-trivial and time-consuming; the
pace of new vulnerability disclosures means the analysis is never really done. Between
2020 and 2026, the NVD accumulated over 198{,}000 CVE records, with over 44{,}000 in
2025 alone; a pace that makes any purely manual process unsustainable.

The core obstacle is a \textit{semantic gap}. On one side sit structured artifacts:
SysML models that describe what software and hardware components exist, Common Platform
Enumeration (CPE) identifiers that name those components precisely, and CVE records
that catalog known flaws. On the other side sit natural-language artifacts: CVE
descriptions written by human analysts, ATT\&CK technique definitions written to
describe adversary behavior, and NIST control narratives written to describe
safeguards. Connecting these two sides requires reading comprehension that rule-based
systems cannot provide. 

Prior work has addressed individual pieces of this problem. Transformer-based
classifiers such as SecRoBERTa can map CVE descriptions to ATT\&CK techniques with
reasonable accuracy~\cite{branescu2024}. Semantic methods such as SMET use structured
role labeling to extract adversary intent from vulnerability text~\cite{abdeen2023,abdeen2024}.
Model-based engineering toolkits such as CEMT can represent threat traceability
explicitly~\cite{fowler2024}. No existing system, however, integrates all three needs:
grounding predictions in the actual system architecture, mapping vulnerabilities to
techniques with contextual awareness, and translating those techniques to recommended
controls. SMSI is designed to fill that gap.

This paper makes four contributions. First, we present a complete end-to-end pipeline
from SysML to NIST 800-53 controls, which to our knowledge is the first system to
automate this entire workflow starting from a SysML model of a CPS gateway.
Second, we introduce a retrieval-based CVE-to-ATT\&CK mapping stage that combines a
TF--IDF lexical baseline, several dense encoders (SecureBERT, ATTACK-BERT, MiniLM),
and KEV-supervised fine-tuning, with diagnostics that quantify where the models agree
or disagree.\cite{grigorescu2022,abdeen2024,host2025}
Third, we evaluate a zero-shot LLM approach (Gemma-4 26B) against both the
retrieval models and a supervised SecureBERT+ classifier, showing that LLMs
achieve substantially higher hit rates on KEV and SMET benchmarks despite having
no task-specific training.
Fourth, we show how ATT\&CK-to-NIST recommendations can be framed as a ranking problem
driven primarily by the CTID ATT\&CK$\to$800-53 crosswalk, with TF--IDF and dense
models providing auxiliary scores and with CVSS severity used to prioritize the result
set for a concrete system of interest.

\section{Literature Review}
\label{sec:litreview}

The problem of automating threat modeling sits at the intersection of three research
threads: NLP-based vulnerability mapping, neuro-symbolic AI in cybersecurity, and
model-based systems engineering (MBSE). We review each in turn, identify their gaps,
and explain where SMSI fits. Table~\ref{tab:sota} summarizes the key approaches.

\subsection{NLP-Based CVE-to-ATT\&CK Mapping}

The foundational work in this area is CVE2ATT\&CK by Grigorescu et al.~\cite{grigorescu2022},
which framed the CVE-to-technique problem as multi-label text classification and
fine-tuned SciBERT on 1,813 labeled pairs, achieving a weighted F1 of 47.84\%.
The main bottleneck was severe class imbalance: ATT\&CK has over 200 techniques,
but most labeled CVEs cluster around a handful of high-frequency ones. 
Branescu et al.~\cite{branescu2024}
expanded the training set to nearly 10,000 entries and found that domain-adapted
encoders matter; SecRoBERTa hit 78.88\% F1 where generic BERT struggled.
Li et al.~\cite{li2024} improved further by operating at sentence granularity rather than
document level and by introducing a relation-based post-processor that enforces
ATT\&CK's own hierarchy. All three approaches treat CVEs in isolation; there is no
notion of what system the CVE applies to. 

Unsupervised and weakly supervised methods reduce the dependence on labeled data.
SMET~\cite{abdeen2023,abdeen2024}
uses Semantic Role Labeling (SRL) to extract subject-verb-object triples from CVE
descriptions and trains a Siamese network to match these triples to ATT\&CK technique
descriptions, achieving Recall@5 of 67.71\% without any labeled CVE-technique pairs.
Recent benchmarks show that hybrid LLM-based systems such as TRIAGE can reach higher
ranked-retrieval scores on the KEV mapping dataset by encoding MITRE's CVE Mapping
Methodology into structured prompts and in-context examples, but still depend on
expert-labeled KEV data and careful governance~\cite{host2025}. 
Rafiey and Namadchian~\cite{rafiey2024} show that large reasoning models such as Gemini
2.5 Pro and OpenAI o1 can reach F1 scores around 60\% in zero-shot settings; reliability
remains a concern due to hallucination, especially when the mapping must be auditable. 

\subsection{Neuro-Symbolic AI in Cybersecurity}

Purely statistical approaches cannot enforce the structural constraints of
cybersecurity. A BERT model might predict a firmware corruption technique for a
cloud-based microservice, which is physically impossible; but the model has no way to
know that. MITREtrieval~\cite{huang2024} fuses a SBERT-based deep learning model with
a cybersecurity ontology (COMAT) using a level-voting algorithm, so that
ontology-derived inferences can validate or override neural predictions. On sparse
datasets, this fusion achieved an F$_2$-score of 69\%, substantially outperforming either
component alone. Graph-based approaches such as AttacKG~\cite{li2022attackg} and 
BRON~\cite{hemberg2021} link CVEs, CWEs, CAPECs, and ATT\&CK entities into a
traversable graph, but these rely on static databases and struggle with new
vulnerability disclosures. 

\subsection{Model-Based Systems Engineering}

The CEMT toolkit~\cite{fowler2024} extends SysML with cybersecurity concepts; misuse
cases, threat vectors, and control mappings; and provides explicit traceability from
system components to NIST 800-53 safeguards. All of CEMT's threat data is entered
manually, however. It provides the scaffolding for traceable threat modeling but no
automation for the analytical steps. SMSI can be thought of as an attempt to automate
the input side of CEMT by wiring a SysML model into a CVE$\to$ATT\&CK$\to$controls
pipeline.

\subsection{Mapping to NIST 800-53}

Beyond CVE-to-ATT\&CK, several authors explore partial automation of control selection.
Sahu and Speretta~\cite{sahu2024} apply TF--IDF and cosine similarity to the text of
NIST 800-53 control families, showing that statistical word analysis can help
semi-automatically suggest which families are most relevant for a given organization's
security documentation. 
Their work confirms that text similarity can meaningfully support, but not replace, human control tailoring. CTID's ATT\&CK-to-CVE impact mapping and ATT\&CK-to-800-53
crosswalk~\cite{nist80053} provide a curated reference that we treat as ground truth in
our ATT\&CK-to-control experiments. 

\subsection{Research Gap}

The literature shows three mature but isolated islands: NLP models for CVE-to-ATT\&CK
mapping, knowledge graphs for structural reasoning, and MBSE frameworks for
traceability. None of these islands are connected to one another, and none starts from
a system architecture model or ends with ranked security controls. SMSI is designed to
bridge all three, while explicitly treating all CVE-to-ATT\&CK inferences as
hypotheses that must be validated against architecture and mission context. 

\begin{table*}[!t]
\renewcommand{\arraystretch}{1.3}
\caption{Comparative Summary of Related Approaches}
\label{tab:sota}
\centering
\begin{tabular}{lllll}
\toprule
\textbf{Approach} & \textbf{Input} & \textbf{Output} & \textbf{System Context} & \textbf{Controls} \\
\midrule
CVE2ATT\&CK / SecRoBERTa~\cite{grigorescu2022,branescu2024} & CVE text & ATT\&CK techniques & None & No \\
SMET~\cite{abdeen2023,abdeen2024} & CVE / CTI text & Ranked ATT\&CK & None & No \\
MITREtrieval~\cite{huang2024} & CTI reports & ATT\&CK techniques & Low & No \\
Generative LLMs~\cite{rafiey2024,host2025} & CVE / prompts & ATT\&CK + summaries & Low--Medium & Generic \\
CEMT~\cite{fowler2024} & SysML model & Risk diagrams + controls & High (manual) & Yes (manual) \\
\textbf{SMSI (this work)} & SysML + CVE feeds & ATT\&CK $\to$ NIST 800-53 & High & Automated (decision support) \\
\bottomrule
\end{tabular}
\end{table*}

\section{Case Study: MedGateway}
\label{sec:casestudy}

To validate the pipeline end-to-end we designed a synthetic but representative system
called \textit{MedGateway}: a healthcare IoT gateway architecture that connects bedside
medical devices to a cloud analytics platform. Healthcare CPS was chosen because
security failures here have direct patient safety consequences, regulatory implications,
and a well-documented vulnerability history. MedGateway is not a real deployed system,
but it is deliberately structured to reflect real healthcare IoT deployments. 

MedGateway consists of nine components organized across four architecture layers,
as specified in the SysML model (\texttt{SOI/MedGateway\_ReferenceArchitecture.sysml.xml}):

\begin{itemize}
    \item \textbf{Application layer:} ClinicalDashboard\_WebApp (Spring Framework 5.3.18)
    \item \textbf{Middleware:} AuditLog\_Service (Apache Log4j 2.14.1), DeviceTelemetry\_Broker (Eclipse Mosquitto 2.0.14)
    \item \textbf{Data layer:} PatientData\_Store (PostgreSQL 14.2)
    \item \textbf{Infrastructure:} APIGateway\_ReverseProxy (Nginx 1.21.6), TLS\_CryptoEngine (OpenSSL 3.0.2), Clinician\_AuthService (Keycloak 18.0.0), Container\_Runtime (Docker Engine 20.10.14), EdgeHost\_OS (Ubuntu 22.04)
\end{itemize}

All components include explicit CPE 2.3 tags in the SysML model, enabling deterministic
NVD matching. Three trust boundaries partition the architecture: IoT~LAN,
Edge~Processing, and Cloud~DMZ.

Log4Shell (CVE-2021-44228, CVSS 10.0) serves as the running example throughout this
paper. It is one of the most severe vulnerabilities ever recorded, it affects Log4j
directly, and its ATT\&CK technique mappings are well-established, making it an ideal
end-to-end test case. 

\section{Datasets}
\label{sec:datasets}

SMSI draws on four publicly available datasets. No proprietary or sensitive data was
used at any stage.

\subsection{National Vulnerability Database}

The NVD is NIST's repository of CVE records~\cite{nvd}. We downloaded bulk JSON feeds
for CVEs published between January 2020 and February 2026.
Our NVD snapshot contains 198{,}783 total vulnerability records across seven annual
files, of which 191{,}740 have usable English-language descriptions
(Table~\ref{tab:nvd_inventory}).
For MedGateway's nine components, CPE-based matching against these feeds returned
199 unique CVEs, distributed as shown in Table~\ref{tab:component_cves}.

\begin{table}[!t]
\renewcommand{\arraystretch}{1.15}
\caption{NVD CVE Inventory (2020--2026)}
\label{tab:nvd_inventory}
\centering
\begin{tabular}{lrr}
\toprule
\textbf{Year} & \textbf{Total CVEs} & \textbf{Usable Descriptions} \\
\midrule
2020 & 21{,}013 & 19{,}340 \\
2021 & 23{,}357 & 22{,}512 \\
2022 & 27{,}468 & 26{,}372 \\
2023 & 31{,}167 & 30{,}551 \\
2024 & 39{,}047 & 38{,}284 \\
2025 & 44{,}010 & 42{,}263 \\
2026 & 12{,}721 & 12{,}418 \\
\midrule
\textbf{Total} & \textbf{198{,}783} & \textbf{191{,}740} \\
\bottomrule
\end{tabular}
\end{table}

\begin{table}[!t]
\renewcommand{\arraystretch}{1.15}
\caption{MedGateway Component CVE Counts}
\label{tab:component_cves}
\centering
\begin{tabular}{lr}
\toprule
\textbf{Component} & \textbf{CVEs} \\
\midrule
EdgeHost\_OS (Ubuntu 22.04) & 65 \\
TLS\_CryptoEngine (OpenSSL 3.0.2) & 40 \\
Clinician\_AuthService (Keycloak 18.0.0) & 27 \\
PatientData\_Store (PostgreSQL 14.2) & 25 \\
Container\_Runtime (Docker 20.10.14) & 17 \\
ClinicalDashboard\_WebApp (Spring 5.3.18) & 10 \\
DeviceTelemetry\_Broker (Mosquitto 2.0.14) & 6 \\
AuditLog\_Service (Log4j 2.14.1) & 5 \\
APIGateway\_ReverseProxy (Nginx 1.21.6) & 4 \\
\midrule
\textbf{Total unique CVEs} & \textbf{199} \\
\bottomrule
\end{tabular}
\end{table}

\subsection{CWE-CAPEC-ATT\&CK Knowledge Graph}

MITRE publishes a chain of mappings linking Common Weakness Enumeration (CWE)
identifiers to Common Attack Pattern Enumeration and Classification (CAPEC) entries,
and from CAPEC entries to ATT\&CK techniques. We used these transitive mappings in
Phase~1 of the supervised classifier pipeline to generate weakly labeled training
data for the SecureBERT+ multi-label model.

\subsection{CISA KEV and CTID Mappings}

The CISA Known Exploited Vulnerabilities (KEV) catalog, cross-referenced with the
CTID ATT\&CK mapping methodology, provides expert-curated (CVE, ATT\&CK technique)
pairs. We use the KEV mapping with 806 mapping objects covering 419 unique CVEs
(ATT\&CK v15.1 alignment) as evaluation ground truth for both the retrieval models
and the LLM approach. The CTID ATT\&CK-to-NIST~800-53 crosswalk covers 470
ATT\&CK techniques and serves as ground truth for the control recommendation stage.

\subsection{NIST 800-53 Rev.\ 5 Control Catalog}

The official NIST 800-53 Rev.\ 5 control catalog~\cite{nist80053} contains
approximately 1{,}000 controls organized into 20 control families (e.g.,
AC---Access Control; SC---System and Communications Protection;
IR---Incident Response). We use this catalog to retrieve full control names and
descriptions for any controls identified by the lookup or similarity-based stages.

\section{Proposed Approach}
\label{sec:approach}

The SMSI pipeline has five sequential stages. We describe each stage here, continuing
to use Log4Shell as the running example.

\subsection{Stage 1: SysML Ingestion and Component Extraction}

The input is a SysML model serialized as XML. The parser walks the XML tree and
extracts each \texttt{sysml:Block} element representing a deployed software or hardware
component. For each block it reads attributes including vendor, product, version, and
an explicit \texttt{cpeHint} tag containing a CPE 2.3 identifier
(e.g., \texttt{cpe:2.3:a:apache:log4j:2.14.1}). When the CPE tag is absent,
a heuristic normalizer constructs a CPE string from the product name and version.
This stage is entirely deterministic; no machine learning is involved.
The output is a component registry of nine SysML block names paired with CPE
identifiers.

\subsection{Stage 2: CPE-to-CVE Lookup via NVD}

For each CPE in the registry, the system queries local NVD JSON feeds over a
configurable time window (2020--2026 in our experiments) and retrieves all matching
CVE records using CPE version-range matching with a fallback to product/version
mention in descriptions. Each record includes a CVE ID, a natural-language
description, a CVSS base score, and optionally a CWE identifier. The result is a
vulnerability set of 199 unique CVEs in which each CVE is linked to one or more SysML
components, preserving full traceability from architecture to flaw.

For Log4j 2.14.1, this stage retrieves CVE-2021-44228 (CVSS 10.0), CVE-2021-45046,
CVE-2021-45105, and two additional entries.

\subsection{Stage 3: CVE-to-ATT\&CK Technique Mapping}

This is the core AI stage and we explore three complementary approaches.

\subsubsection{Approach A: Retrieval-Based Dense Encoders}

We treat CVE-to-ATT\&CK as ranked retrieval rather than classification.
A TF--IDF baseline fits a vectorizer on ATT\&CK technique descriptions, a system-of-interest
(SOI) profile, and a sample of NVD descriptions. For each CVE, cosine similarity
against all techniques yields a ranked list.

We then replace sparse vectors with dense embeddings from three encoders:
SecureBERT (\texttt{ehsanaghaei/SecureBERT}) with HF mean-pooling,
ATTACK-BERT (\texttt{basel/ATTACK-BERT}) via \texttt{SentenceTransformer},
and MiniLM (\texttt{all-MiniLM-L6-v2}) as an unsupervised baseline.
A DeBERTa-v3 NLI cross-encoder reranks MiniLM's top candidates by entailment probability.

To inject supervision, we fine-tune SecureBERT and ATTACK-BERT on KEV (CVE, technique) pairs
using \texttt{MultipleNegativesRankingLoss} with 80/20 CVE-ID-grouped train/test splits.

\subsubsection{Approach B: Supervised Multi-Label Classification}

Separately, we train a supervised SecureBERT+ classifier on a large weakly labeled
dataset constructed via CWE$\to$CAPEC$\to$ATT\&CK transitive mappings and KEV gold labels.
Phase~1 assembles $\sim$191K labeled CVE-technique pairs; Phase~1.5 collapses
sub-techniques to 105 parent classes; Phase~2 cleans text and applies a minimum
support filter; Phase~4 fine-tunes \texttt{ehsanaghaei/SecureBERT\_Plus} with
\texttt{BCEWithLogitsLoss} for multi-label classification; and Phase~5 performs
threshold optimization on the test set.

\subsubsection{Approach C: Zero-Shot LLM Inference}

We evaluate a local LLM (Gemma-4 26B, served via LM~Studio) in a zero-shot setting.
The LLM receives a structured prompt containing the CVE description and the full
ATT\&CK parent technique catalogue (214 techniques with names), and returns a
JSON list of up to 5 ranked technique predictions. We test two modes: \emph{no hint}
(CVE description only) and \emph{DB hint} (with CWE$\to$CAPEC-derived technique
suggestions appended to the prompt). Evaluation uses two benchmarks:
the KEV dataset (419 CVEs) and the SMET academic dataset (302 CVEs).

\subsection{Stage 4: ATT\&CK-to-NIST 800-53 Control Recommendation}

Given the ATT\&CK techniques predicted for a component's CVEs, this stage recommends
NIST 800-53 Rev.\ 5 controls using three mechanisms.

\subsubsection{Direct Lookup via CTID Crosswalk}

The CTID mapping covers 470 ATT\&CK techniques. For any technique in this set, the
associated controls are retrieved directly; this is a deterministic lookup with
precision 1.0 on crosswalk-covered techniques~\cite{nist80053}. 

\subsubsection{TF--IDF and Dense Similarity}

We reuse the TF--IDF vectorizer from Stage~3, now fit on ATT\&CK technique texts,
NIST control texts, and the SOI profile. Dense embeddings from SecureBERT, ATTACK-BERT,
and MiniLM produce additional similarity scores. Fine-tuned variants of SecureBERT and
ATTACK-BERT trained on CTID crosswalk pairs using \texttt{MultipleNegativesRankingLoss}
are also evaluated.

\subsubsection{Hybrid Control Ranking and CVSS Weighting}

A hybrid control score combines the binary crosswalk indicator and TF--IDF similarity:
\[
\text{hybrid\_score} = 0.72 \cdot \text{crosswalk\_score} + 0.28 \cdot \text{tfidf\_score},
\]
where the crosswalk score is 1.0 for pairs appearing in the CTID mapping and 0
otherwise. CVSS scores from Stage~2 propagate into a final priority score:
\[
\text{priority} = \text{hybrid\_mapping\_soi} \times \text{max\_cvss}.
\]

\subsection{Stage 5: Report Generation and Traceability}

For every ranked control, SMSI emits the full trace:
(SysML component $\rightarrow$ CPE $\rightarrow$ CVE $\rightarrow$ ATT\&CK technique
$\rightarrow$ NIST control), along with intermediate similarity scores and CVSS values.

\section{Experimental Methodology}
\label{sec:methodology}

\subsection{CVE-to-ATT\&CK Evaluation: Retrieval Models}

We evaluate retrieval models against a held-out set of 45 KEV CVEs
(\texttt{GroupShuffleSplit}, 20\% test). Each model is evaluated as a ranked retrieval
system using MRR and Hits@$K$ for $K\in\{1,3,5,10\}$.
We compare six approaches: TF--IDF, pretrained SecureBERT, pretrained ATTACK-BERT,
fine-tuned SecureBERT, fine-tuned ATTACK-BERT, and MiniLM (unsupervised).

\subsection{CVE-to-ATT\&CK Evaluation: Supervised Classifier}

\begin{figure}[!t]
\centering
\includegraphics[width=0.92\linewidth]{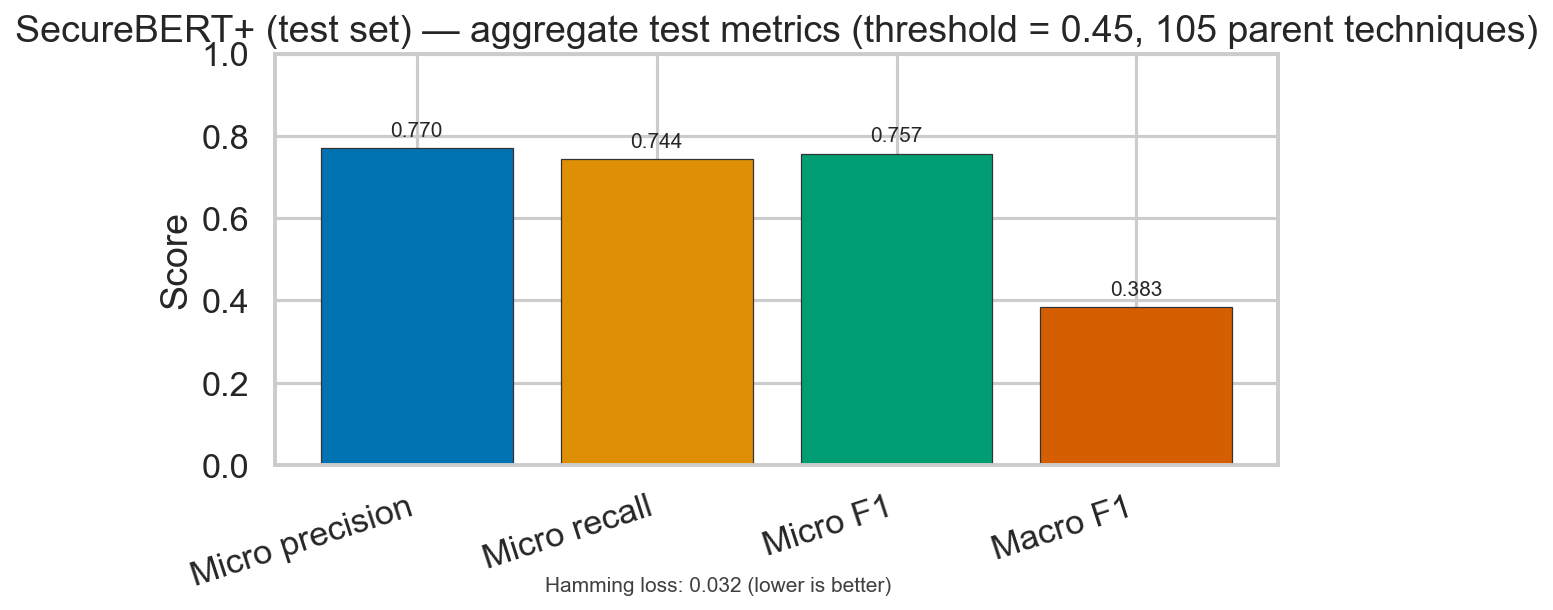}
\caption{SecureBERT+ aggregate test metrics at optimal threshold 0.45 (105 parent techniques). Hamming loss is 0.032 (table).}
\label{fig:securebert_plus_aggregate}
\end{figure}

The SecureBERT+ multi-label classifier is evaluated on a held-out test split using
threshold optimization (sweeping 0.10--0.95 in steps of 0.05), reporting Micro/Macro
F1, precision, recall, Hamming loss, and per-class F1 at the optimal threshold.
Quantitative results are summarized in Section~\ref{sec:results}
(Table~\ref{tab:securebert_classifier}, Figures~\ref{fig:securebert_plus_aggregate}
and~\ref{fig:securebert_class_f1}).

\subsection{CVE-to-ATT\&CK Evaluation: LLM Zero-Shot}

The Gemma-4 26B LLM is evaluated on two complete benchmarks without sampling:
419 KEV CVEs and 302 SMET CVEs. We report Hit Rate@$K$ (fraction of CVEs where at
least one true technique appears in the top-$K$ predictions), true Recall@$K$ (fraction
of all true labels recovered), and classification metrics (weighted F1, macro F1,
LRAP).

\subsection{ATT\&CK-to-NIST Control Evaluation}

The CTID crosswalk serves as ground truth. We split by technique ID into 80\% train
and 20\% test (94 test queries) and evaluate the same model families using MRR
and Hits@$K$.

\section{Results}
\label{sec:results}

\subsection{CVE-to-ATT\&CK: Retrieval Models}

\begin{figure}[!t]
\centering
\includegraphics[width=\linewidth]{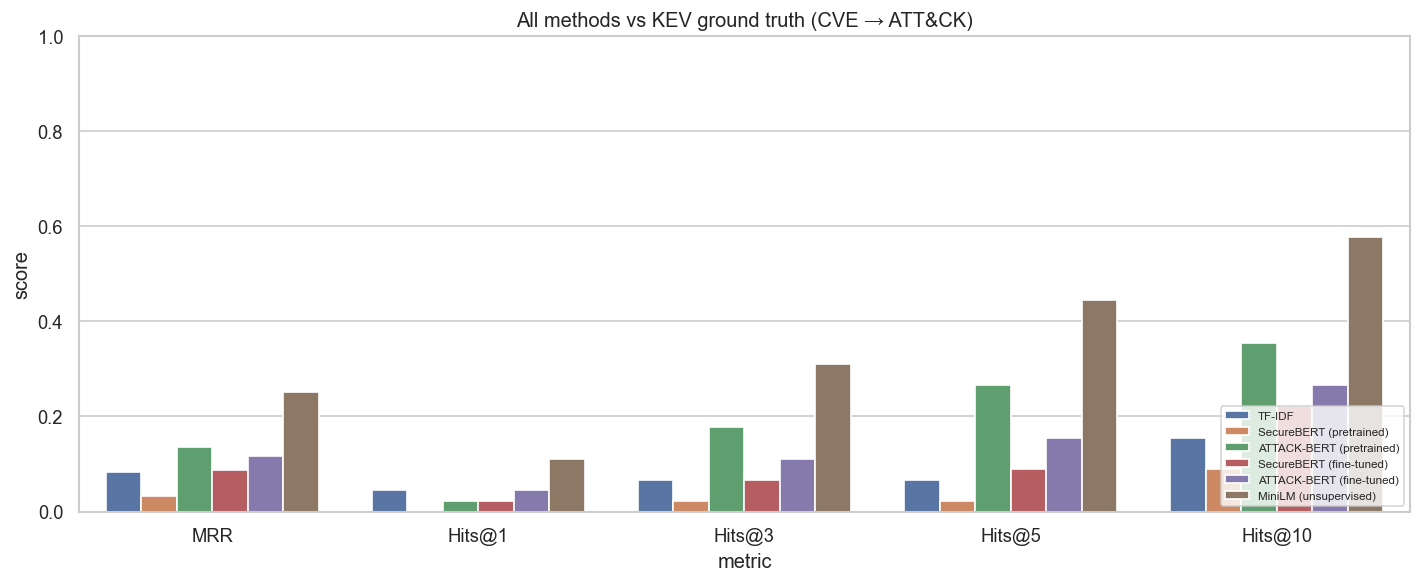}
\caption{CVE-to-ATT\&CK retrieval: all methods vs KEV ground truth ($n=45$ test CVEs). MiniLM dominates all metrics.}
\label{fig:cve_all_methods}
\end{figure}

Table~\ref{tab:cve_retrieval} summarizes the retrieval evaluation on the 45-query KEV
test split. MiniLM (unsupervised) achieves the highest scores across all metrics,
with MRR of 0.252 and Hits@10 of 0.578. Pretrained ATTACK-BERT is the second-best
model (MRR 0.136, Hits@10 0.356). Fine-tuning improves SecureBERT substantially
(MRR from 0.033 to 0.087) but unexpectedly degrades ATTACK-BERT (MRR from 0.136
to 0.117). Pretrained SecureBERT performs worst, even below TF--IDF.
Figure~\ref{fig:cve_all_methods} visualizes these results.

\begin{table}[!t]
\renewcommand{\arraystretch}{1.15}
\caption{CVE-to-ATT\&CK Retrieval (KEV Test, $n=45$)}
\label{tab:cve_retrieval}
\centering
\begin{tabular}{lcccc}
\toprule
\textbf{Model} & \textbf{MRR} & \textbf{H@1} & \textbf{H@5} & \textbf{H@10} \\
\midrule
TF--IDF & 0.082 & 0.044 & 0.067 & 0.156 \\
SecureBERT (pretrained) & 0.033 & 0.000 & 0.022 & 0.089 \\
SecureBERT (fine-tuned) & 0.087 & 0.022 & 0.089 & 0.222 \\
ATTACK-BERT (pretrained) & 0.136 & 0.022 & 0.267 & 0.356 \\
ATTACK-BERT (fine-tuned) & 0.117 & 0.044 & 0.156 & 0.267 \\
MiniLM (unsupervised) & \textbf{0.252} & \textbf{0.111} & \textbf{0.444} & \textbf{0.578} \\
\bottomrule
\end{tabular}
\end{table}

\begin{figure}[!t]
\centering
\includegraphics[width=\linewidth]{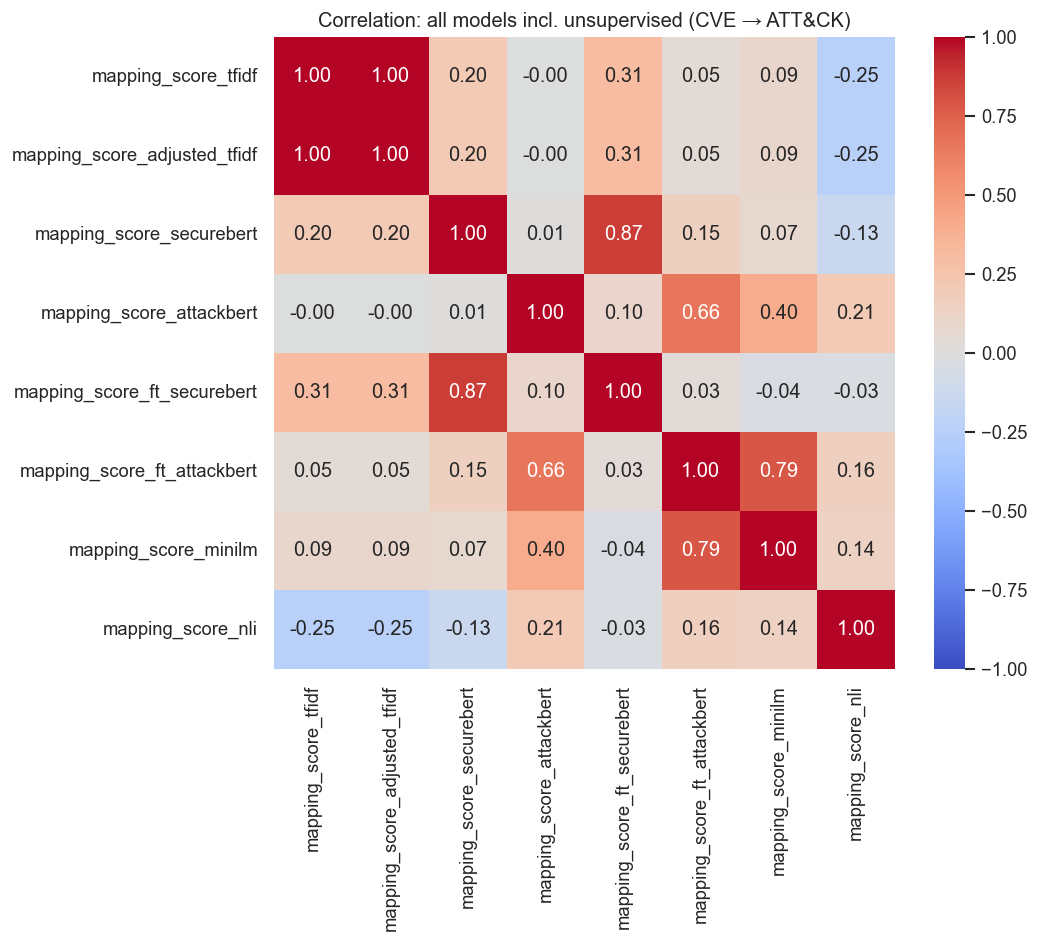}
\caption{Pearson correlation between all CVE-to-ATT\&CK model scores. TF--IDF and dense models capture largely independent signals.}
\label{fig:cve_correlation}
\end{figure}

The correlation heatmap (Figure~\ref{fig:cve_correlation}) reveals that models capture
fundamentally different signals. TF--IDF and pretrained ATTACK-BERT are uncorrelated
($r \approx 0.00$), while fine-tuned ATTACK-BERT and MiniLM are strongly correlated
($r = 0.79$). Fine-tuned SecureBERT correlates highly with its pretrained version
($r = 0.87$) but not with other dense models.

\subsection{CVE-to-ATT\&CK: Supervised Classifier}

The SecureBERT+ multi-label classifier, trained on the full CWE/CAPEC/KEV dataset and
evaluated on 105 parent technique classes, achieves the results shown in
Table~\ref{tab:securebert_classifier}. The optimal threshold is 0.45.

\begin{table}[!t]
\renewcommand{\arraystretch}{1.15}
\caption{SecureBERT+ Multi-Label Classifier (Parent Techniques, 105 Classes)}
\label{tab:securebert_classifier}
\centering
\begin{tabular}{lc}
\toprule
\textbf{Metric} & \textbf{Score} \\
\midrule
Micro Precision & 0.770 \\
Micro Recall & 0.744 \\
Micro F1 & \textbf{0.757} \\
Macro F1 & 0.383 \\
Hamming Loss & 0.032 \\
Optimal Threshold & 0.45 \\
\bottomrule
\end{tabular}
\end{table}

\begin{figure}[!t]
\centering
\includegraphics[width=\linewidth]{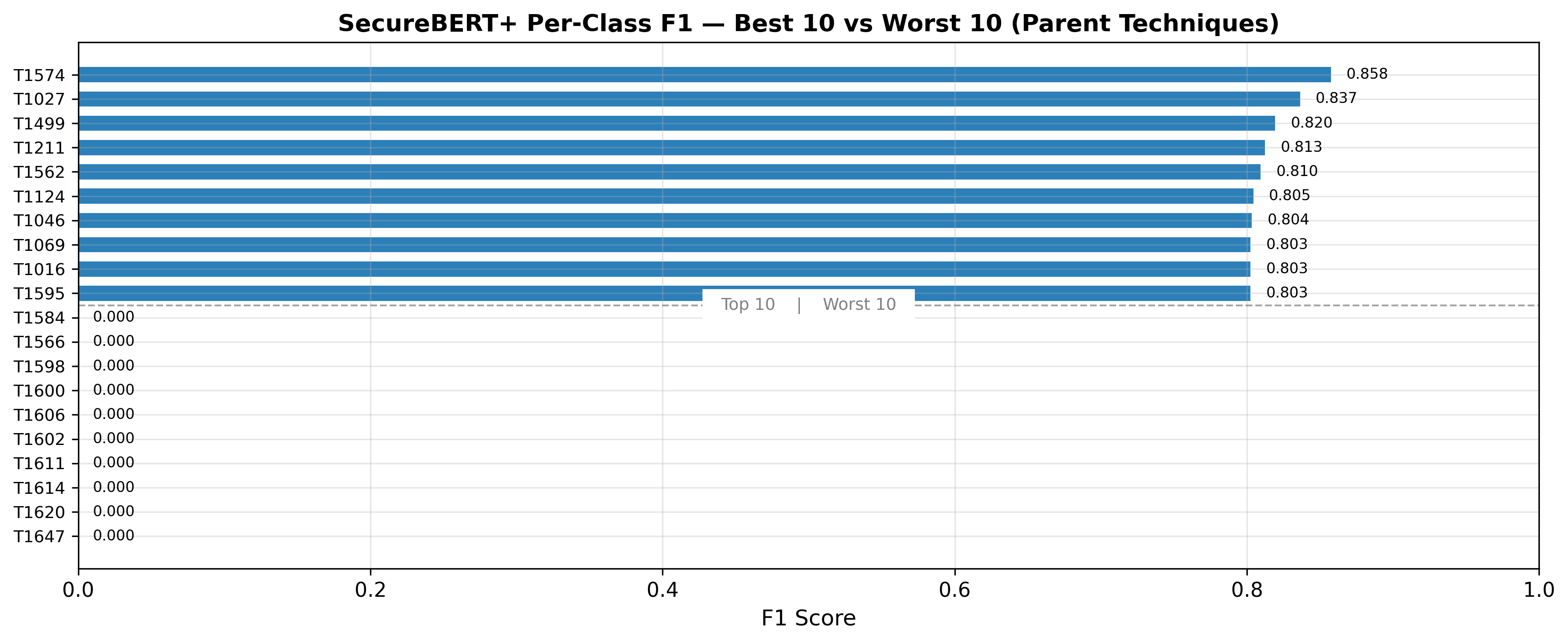}
\caption{SecureBERT+ per-class F1: best 10 vs worst 10 parent techniques. The long tail of zero-scoring classes drives the low Macro F1.}
\label{fig:securebert_class_f1}
\end{figure}

While the Micro F1 of 0.757 is competitive with prior work, the low Macro F1 (0.383)
reveals severe class imbalance: the top-performing techniques (e.g., T1574 at F1=0.858,
T1027 at F1=0.837) have thousands of training samples, while 10+ techniques score
F1=0.000 due to insufficient support
(Figure~\ref{fig:securebert_class_f1}).

\subsection{CVE-to-ATT\&CK: Zero-Shot LLM}

\begin{figure}[!t]
\centering
\includegraphics[width=\linewidth]{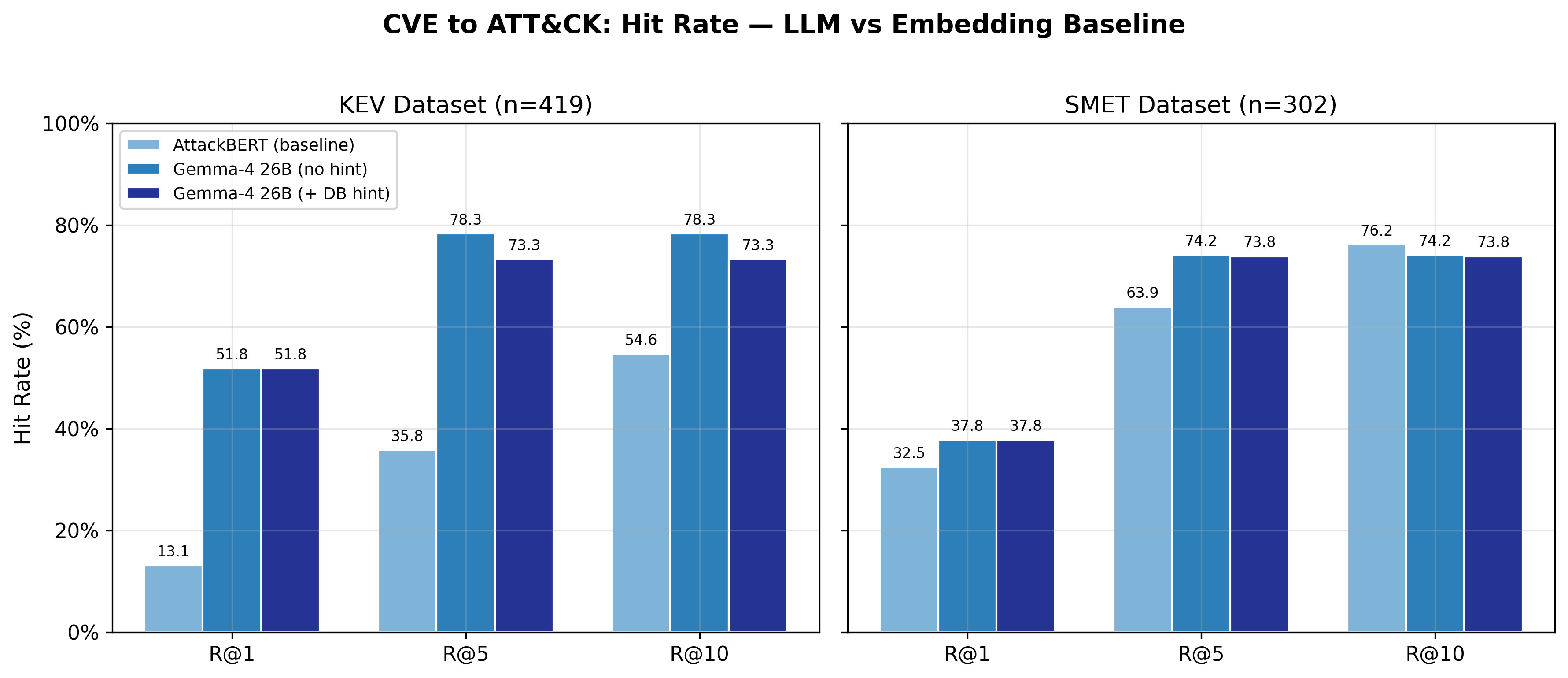}
\caption{Hit Rate comparison: Gemma-4 26B vs ATTACK-BERT embedding baseline on KEV (419 CVEs) and SMET (302 CVEs).}
\label{fig:llm_hitrate}
\end{figure}

The Gemma-4 26B LLM achieves substantially higher hit rates than all retrieval models
(Table~\ref{tab:llm_results}), with Hit Rate@1 of 51.8\% on KEV (419 CVEs) and
37.8\% on SMET (302 CVEs). On the KEV dataset, this represents a $\sim$4$\times$
improvement over the best embedding baseline (ATTACK-BERT, 13.1\%).
However, true recall metrics are more modest (R@5 of 35.4\% on KEV, 62.5\%
on SMET), and the LLM's weighted F1 scores (0.282 on KEV, 0.411 on SMET) fall below
the supervised SecureBERT+ classifier. Adding CWE/CAPEC-derived database hints did
not improve results, slightly degrading R@5 on both datasets.
Figure~\ref{fig:llm_hitrate} compares the LLM against the ATTACK-BERT baseline.

\begin{table}[!t]
\renewcommand{\arraystretch}{1.15}
\caption{Gemma-4 26B Zero-Shot LLM Evaluation}
\label{tab:llm_results}
\centering
\begin{tabular}{lcccc}
\toprule
\textbf{Dataset / Mode} & \textbf{HR@1} & \textbf{HR@5} & \textbf{R@5} & \textbf{W-F1} \\
\midrule
KEV, no hint ($n$=419) & 51.8 & 78.3 & 35.4 & 0.282 \\
KEV, +DB hint ($n$=419) & 51.8 & 73.3 & --- & --- \\
SMET, no hint ($n$=302) & 37.8 & 74.2 & 62.5 & 0.411 \\
SMET, +DB hint ($n$=302) & 37.8 & 73.8 & --- & --- \\
\bottomrule
\end{tabular}
\end{table}

\begin{figure}[!t]
\centering
\includegraphics[width=\linewidth]{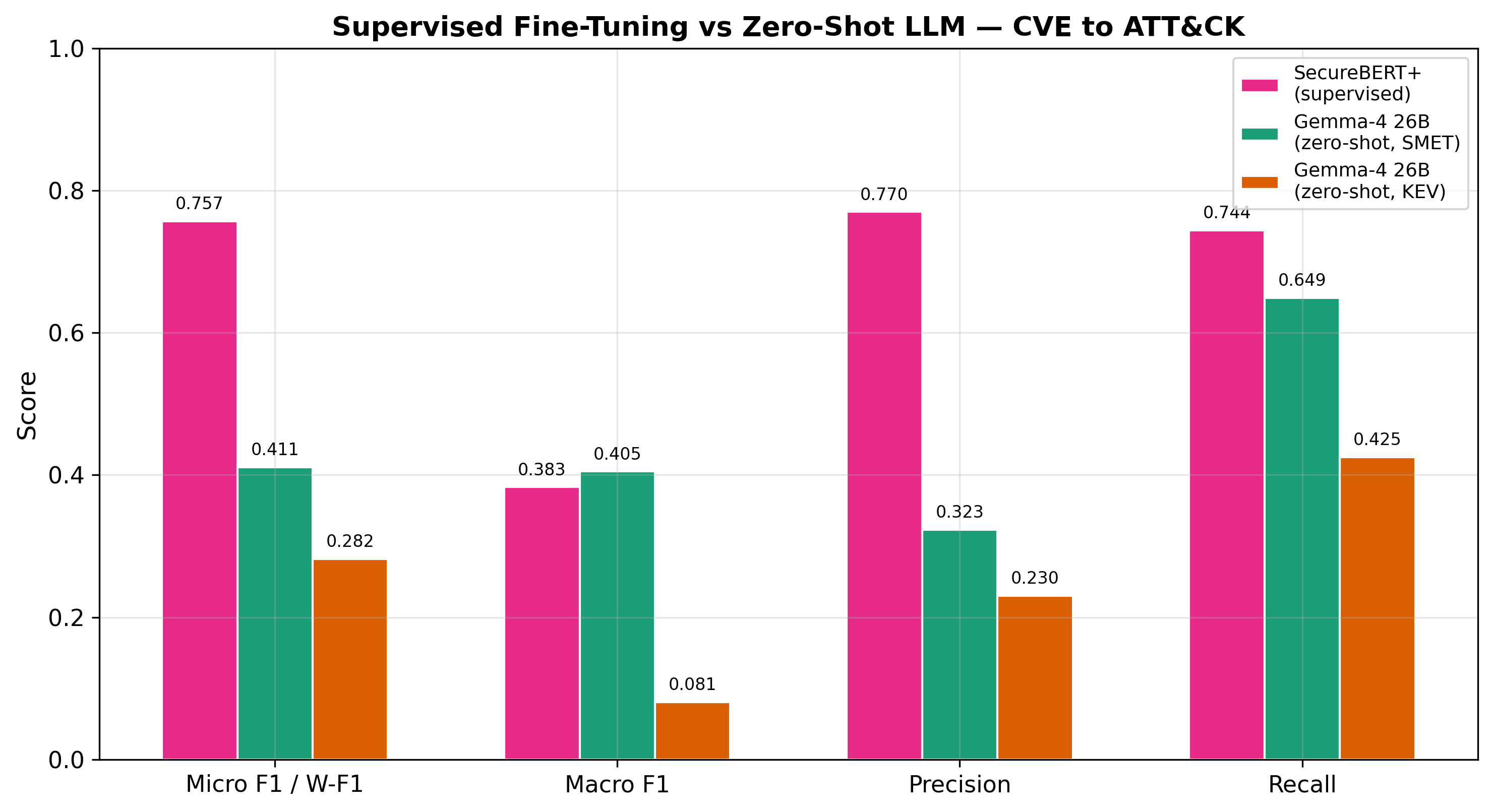}
\caption{Supervised SecureBERT+ fine-tuning vs zero-shot Gemma-4 26B. The supervised model leads on precision and F1; the LLM is more competitive on recall and rare-class coverage.}
\label{fig:securebert_vs_llm}
\end{figure}

Figure~\ref{fig:securebert_vs_llm} provides a cross-paradigm comparison of the
supervised classifier against the zero-shot LLM.
The supervised SecureBERT+ dominates on precision-oriented metrics (Micro~F1 0.757
vs W-F1 0.411), while the LLM shows stronger performance on macro F1 for the SMET
dataset (0.405 vs 0.383), suggesting it handles rare technique classes more uniformly.

\subsection{ATT\&CK-to-NIST Control Recommendation}

Table~\ref{tab:atk_controls} shows the ATT\&CK-to-controls retrieval results on
94 test technique queries. Pretrained SecureBERT achieves the highest scores by a
wide margin (MRR 0.582, Hits@10 0.798), substantially outperforming all other models
including its own fine-tuned variant (MRR 0.239). This reversal from the CVE-to-ATT\&CK
results suggests that SecureBERT's pretrained embeddings already encode strong
alignment between security concept descriptions and control language, and that
fine-tuning on the relatively small crosswalk dataset introduces overfitting.

\begin{table}[!t]
\renewcommand{\arraystretch}{1.15}
\caption{ATT\&CK-to-NIST Control Retrieval (Crosswalk Test, $n=94$)}
\label{tab:atk_controls}
\centering
\begin{tabular}{lcccc}
\toprule
\textbf{Model} & \textbf{MRR} & \textbf{H@1} & \textbf{H@5} & \textbf{H@10} \\
\midrule
TF--IDF & 0.179 & 0.032 & 0.340 & 0.404 \\
SecureBERT (pretrained) & \textbf{0.582} & \textbf{0.447} & \textbf{0.755} & \textbf{0.798} \\
SecureBERT (fine-tuned) & 0.239 & 0.128 & 0.372 & 0.500 \\
ATTACK-BERT (pretrained) & 0.113 & 0.043 & 0.128 & 0.266 \\
ATTACK-BERT (fine-tuned) & 0.195 & 0.085 & 0.287 & 0.436 \\
MiniLM (unsupervised) & 0.124 & 0.043 & 0.181 & 0.277 \\
\bottomrule
\end{tabular}
\end{table}

\begin{figure}[!t]
\centering
\includegraphics[width=\linewidth]{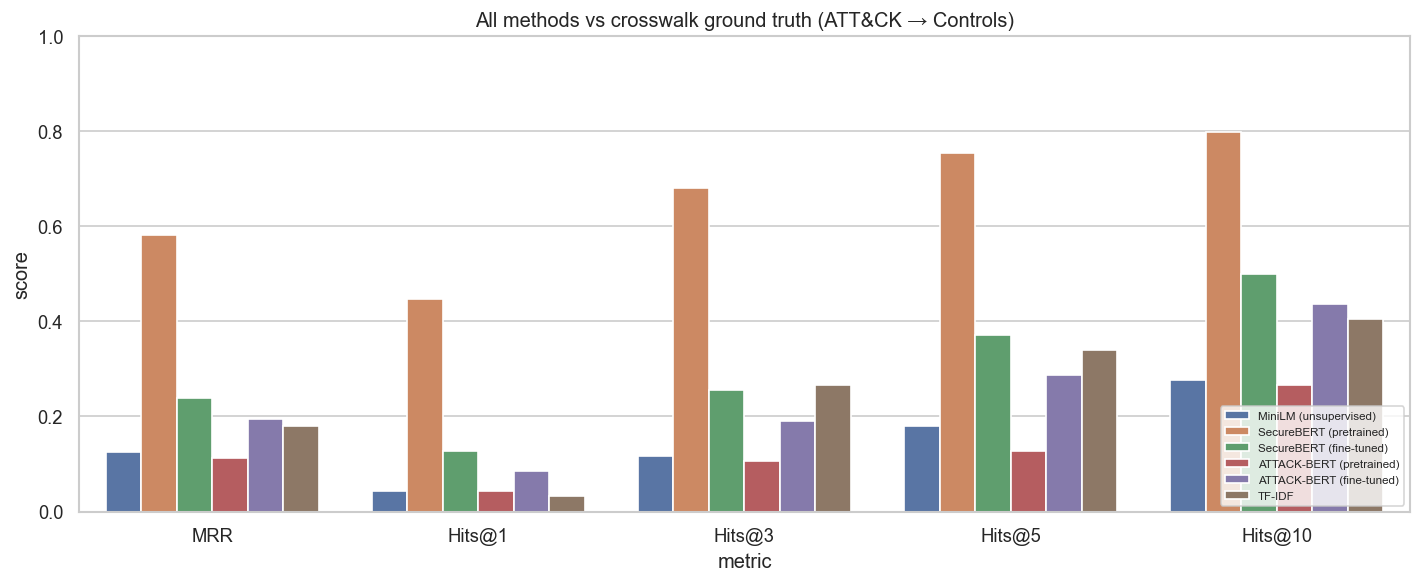}
\caption{ATT\&CK-to-controls retrieval: all methods vs CTID crosswalk ground truth ($n=94$). Pretrained SecureBERT dominates.}
\label{fig:atk_controls}
\end{figure}

\begin{figure}[!t]
\centering
\includegraphics[width=\linewidth]{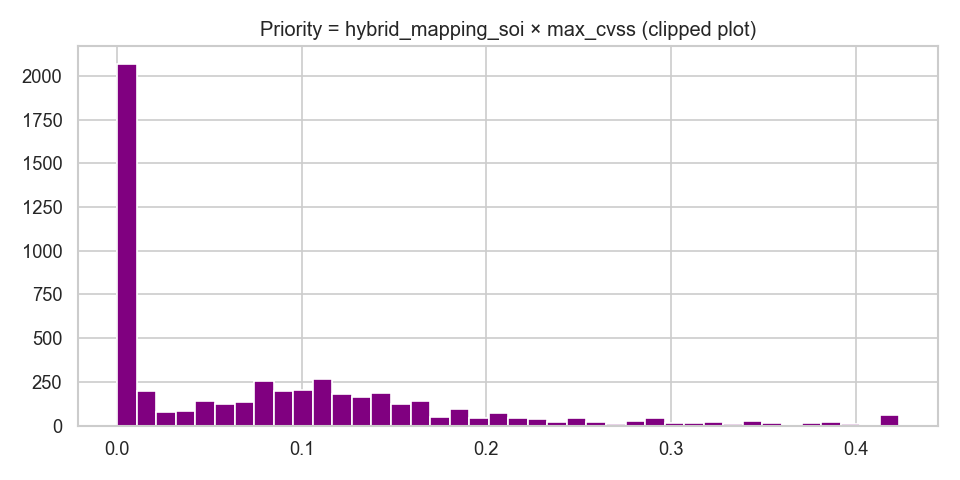}
\caption{Priority distribution for technique-control pairs
(\texttt{priority = hybrid\_mapping\_soi} $\times$ \texttt{max\_cvss}); clipped at the 99th percentile.}
\label{fig:priority_hist}
\end{figure}

The CVSS-weighted priority distribution (Figure~\ref{fig:priority_hist}) is
heavy-tailed: most technique-control pairs cluster near zero, while a small fraction
dominate the upper tail. This is desirable for triage because it concentrates analyst
attention on a manageable subset of high-impact recommendations.

\subsection{End-to-End Pipeline: MedGateway}

Tracing Log4Shell through the full pipeline illustrates how the stages connect.
The SysML parser identifies the AuditLog\_Service block with
\texttt{cpe:2.3:a:apache:log4j:2.14.1}. The NVD lookup returns 5 CVEs including
CVE-2021-44228 with CVSS 10.0. The CVE-to-ATT\&CK stage assigns high scores to
techniques such as T1190 (Exploit Public-Facing Application) and T1059 (Command and
Scripting Interpreter) across multiple models. The CTID crosswalk links these to
controls including AC-3 (Access Enforcement), SI-10 (Information Input Validation),
SC-7 (Boundary Protection), and SI-3 (Malicious Code Protection). The CVSS-weighted
hybrid score pushes Log4Shell-related pairs toward the head of the global priority
list.

\section{Discussion}
\label{sec:discussion}

\subsection{Why MiniLM Outperformed Domain-Specific Models}

The most surprising CVE-to-ATT\&CK result is that MiniLM (a generic 22M-parameter
sentence encoder with no security-specific pretraining) outperforms both frozen and
fine-tuned domain-specific models. We hypothesize two contributing factors.
First, MiniLM was trained with a contrastive objective on diverse sentence pairs,
producing an embedding space well-suited for cross-domain semantic similarity. Security
domain pretraining (SecureBERT) or ATT\&CK-specific training (ATTACK-BERT) may actually
narrow the embedding space in ways that hurt cross-document retrieval between CVE
descriptions and technique definitions, which use fundamentally different vocabulary.
Second, the KEV fine-tuning set is small (fewer than 200 training CVEs), making
overfitting likely for the larger domain models.

\subsection{Supervised vs Zero-Shot: Complementary Strengths}

The three CVE-to-ATT\&CK approaches occupy different points in the precision-coverage
tradeoff. The supervised SecureBERT+ classifier achieves the highest Micro F1 (0.757)
but only on techniques with sufficient training data; its Macro F1 of 0.383 shows it
effectively ignores rare classes. The Gemma-4 LLM achieves lower overall F1 but
distributes its predictions more uniformly across the technique space (SMET Macro F1
of 0.405), suggesting that LLMs are better at rare-technique coverage. The retrieval
models (MiniLM) provide ranked lists rather than binary predictions, making them
suitable for shortlisting candidates that analysts can review. A production system
could ensemble all three: use the classifier for confident high-frequency predictions,
the LLM for rare techniques, and retrieval scores for ranking.

\subsection{SecureBERT's Surprising Control-Retrieval Dominance}

In the ATT\&CK-to-controls stage, pretrained SecureBERT (MRR 0.582) dramatically
outperforms all other models including MiniLM (MRR 0.124). This is the opposite of
the CVE-to-ATT\&CK pattern, where SecureBERT performed worst. The explanation lies in
the nature of the text: ATT\&CK technique descriptions and NIST control descriptions
both use formal security terminology, which is exactly what SecureBERT's pretraining
corpus covers. CVE descriptions, by contrast, mix security language with product-specific
jargon (e.g., JNDI lookups, Spring DataBinder), creating a vocabulary mismatch that
SecureBERT's pretraining does not resolve.

\subsection{Limitations}

The NVD is queried as a static snapshot; new vulnerabilities disclosed after the query
date are not captured. The KEV fine-tuning split contains only 45 test CVEs, limiting
statistical power. The LLM evaluation depends on a local model server, making exact
reproducibility hardware-dependent. The CWE/CAPEC knowledge graph is used for training
data generation but not yet for inference-time reasoning. All models operate on
English-language artifacts and the Enterprise ATT\&CK profile; extending to ICS/OT
ATT\&CK remains future work. CPE matching can fail for vague SysML descriptions.

\section{Conclusion}
\label{sec:conclusion}

SMSI demonstrates that the end-to-end threat modeling workflow, from system
architecture to ranked security controls, can be meaningfully automated for a concrete
SysML case study with nine components and 199 CVEs. The key insight is that the problem
has two distinct sub-problems that benefit from different strategies: structured data
lookups (SysML to CPE to CVE; ATT\&CK to NIST for crosswalk-covered techniques) are
best handled deterministically, while semantic reasoning over natural language (CVE to
ATT\&CK; ATT\&CK to NIST for uncovered techniques) benefits from neural models
grounded by symbolic constraints and curated mappings.

Our experiments reveal that no single approach dominates CVE-to-ATT\&CK mapping.
The supervised SecureBERT+ classifier achieves the best Micro F1 (0.757) but suffers
from severe class imbalance. The zero-shot Gemma-4 LLM achieves dramatically higher
hit rates (51.8\% HR@1 vs $<$15\% for embedding models) and better rare-class coverage.
The unsupervised MiniLM provides the best retrieval-based ranking (MRR 0.252), offering
a strong baseline that requires no labeled data. For ATT\&CK-to-controls, pretrained
SecureBERT embeddings provide surprisingly effective retrieval (MRR 0.582), and the
CTID crosswalk remains the most reliable backbone signal.

From a practical standpoint, SMSI reduces what would otherwise require days of manual
expert analysis to an automated pipeline producing fully traceable recommendations---
provided they are reviewed by human analysts rather than treated as an oracle.

\section{Future Work}
\label{sec:futurework}

The most impactful near-term improvement would be making the pipeline explicitly
\emph{architecture-conditioned}: incorporating network exposure, component criticality,
and data sensitivity into the risk weighting formula so that a Log4j instance on the
public internet triggers different prioritization than one on an isolated subnet.
Real-time NVD integration would enable continuous monitoring. The CVE-to-ATT\&CK stage
could incorporate the CWE/CAPEC knowledge graph at inference time, combining
deterministic graph traversal with neural retrieval in a principled neuro-symbolic
fashion~\cite{abdeen2023,hemberg2021}. An ensemble of the supervised classifier,
LLM, and retrieval models could be calibrated to optimize precision-recall tradeoffs
for specific deployment contexts. Finally, a user study with practicing security
analysts would measure whether SMSI's recommendations genuinely reduce manual effort.


\section*{Acknowledgment}
The authors would like to thank Prof.\ Paula Branco for her guidance throughout
the course, and the course TA for their feedback during the progress check meeting.
Ro'Yah also gratefully acknowledges the mentorship and support of
Roderick Fernandes, Senior Special Advisor / Cyber Security Engineer at DND, whose
guidance over the past three years has shaped much of her understanding of this field.




\begin{thebibliography}{18}

\bibitem{grigorescu2022}
O.~Grigorescu, A.~Nica, M.~Dascalu, and R.~Rughinis,
``CVE2ATT\&CK: BERT-based mapping of CVEs to MITRE ATT\&CK techniques,''
\emph{Algorithms}, vol.~15, no.~9, Aug.~2022,
doi: 10.3390/a15090314.

\bibitem{branescu2024}
I.~Branescu, O.~Grigorescu, and M.~Dascalu,
``Automated mapping of common vulnerabilities and exposures to MITRE ATT\&CK tactics,''
\emph{Information}, vol.~15, no.~4, Apr.~2024,
doi: 10.3390/info15040214.

\bibitem{li2024}
L.~Li, C.~Huang, and J.~Chen,
``Automated discovery and mapping ATT\&CK tactics and techniques for unstructured
cyber threat intelligence,''
\emph{Comput. Secur.}, vol.~140, p.~103815, May~2024,
doi: 10.1016/j.cose.2024.103815.

\bibitem{abdeen2023}
B.~Abdeen, E.~Al-Shaer, A.~Singhal, L.~Khan, and K.~Hamlen,
``SMET: Semantic mapping of CVE to ATT\&CK and its application to cybersecurity,''
in \emph{Data and Applications Security and Privacy XXXVII},
V.~Atluri and A.~L.~Ferrara, Eds.
Cham: Springer, 2023, pp.~243--260,
doi: 10.1007/978-3-031-37586-6\_15.

\bibitem{abdeen2024}
B.~Abdeen \emph{et al.},
``SMET: Semantic mapping of CTI reports and CVE to ATT\&CK for advanced threat intelligence,''
\emph{J. Comput. Secur.}, 2024, to appear. [Online]. Available:
\url{https://doi.org/10.3233/JCS-230218}.

\bibitem{kuppa2021}
A.~Kuppa, L.~Aouad, and N.-A.~Le-Khac,
``Linking CVE's to MITRE ATT\&CK techniques,''
in \emph{Proc. 16th Int. Conf. Availability, Reliability and Security (ARES)},
New York, NY, USA: ACM, Aug.~2021, pp.~1--12,
doi: 10.1145/3465481.3465758.

\bibitem{liu2023}
X.~Liu, Y.~Tan, Z.~Xiao, J.~Zhuge, and R.~Zhou,
``Not the end of story: An evaluation of ChatGPT-driven vulnerability description
mappings,''
in \emph{Findings of the Assoc. Comput. Linguist.: ACL 2023},
Toronto, Canada: ACL, Jul.~2023, pp.~3724--3731,
doi: 10.18653/v1/2023.findings-acl.229.

\bibitem{rafiey2024}
P.~Rafiey and A.~Namadchian,
``Mapping vulnerability description to MITRE ATT\&CK framework by LLM,''
May~2024, Research Square,
doi: 10.21203/rs.3.rs-4341401/v1.

\bibitem{huang2024}
Y.-T.~Huang \emph{et al.},
``MITREtrieval: Retrieving MITRE techniques from unstructured threat reports by
fusion of deep learning and ontology,''
\emph{IEEE Trans. Netw. Service Manag.}, vol.~21, no.~4, pp.~4871--4887, Aug.~2024,
doi: 10.1109/TNSM.2024.3401200.

\bibitem{li2022attackg}
Z.~Li, J.~Zeng, Y.~Chen, and Z.~Liang,
``AttacKG: Constructing technique knowledge graph from cyber threat intelligence
reports,''
arXiv:2111.07093, May~2022,
doi: 10.48550/arXiv.2111.07093.

\bibitem{hemberg2021}
E.~Hemberg \emph{et al.},
``Linking threat tactics, techniques, and patterns with defensive weaknesses,
vulnerabilities and affected platform configurations for cyber hunting,''
arXiv:2010.00533, Feb.~2021,
doi: 10.48550/arXiv.2010.00533.

\bibitem{fowler2024}
S.~Fowler, K.~Joiner, and S.~Ma,
``Cyber Evaluation and Management Toolkit (CEMT): Face validity of model-based
cybersecurity decision making,''
\emph{Systems}, vol.~12, no.~7, Jun.~2024,
doi: 10.3390/systems12070238.

\bibitem{host2025}
A.~H\o st, P.~Lison, and L.~Moonen,
``A systematic approach to predict the impact of cybersecurity vulnerabilities using LLMs,''
2025. [Online]. Available: \url{https://arxiv.org/abs/2508.18439}.

\bibitem{sahu2024}
R.~Sahu and M.~Speretta,
``Statistical word analysis to support the semiautomatic implementation of the NIST 800-53 cybersecurity framework,''
in \emph{Proc. ASEE Annu. Conf.}, 2024.

\bibitem{nist80053}
Joint Task Force,
\emph{Security and Privacy Controls for Information Systems and Organizations},
NIST Special Publication 800-53 Rev.~5,
National Institute of Standards and Technology, Dec.~2020,
doi: 10.6028/NIST.SP.800-53r5.

\bibitem{nvd}
National Institute of Standards and Technology,
``National Vulnerability Database,''
[Online]. Available: \url{https://nvd.nist.gov/}.
Accessed: Feb.~21, 2026.

\bibitem{attck}
MITRE Corporation,
``MITRE ATT\&CK,''
[Online]. Available: \url{https://attack.mitre.org/}.
Accessed: Feb.~21, 2026.

\end{thebibliography}
\end{document}